\documentclass[a4paper,12pt]{article}
\usepackage{rsc}
\usepackage{amsmath}    
\usepackage{graphicx}   
\usepackage{verbatim}   
\usepackage{color}      
\usepackage{subfigure}  
\usepackage[linkcolor=blue]{hyperref}   
\usepackage{amssymb}
\usepackage[margin=1.5in]{geometry} 
\usepackage[font=footnotesize,labelfont=bf]{caption} 

\newcommand{\ttss}[1]{\textsuperscript{#1}}

\begin{document}

\begin{center}
{\Large
\textbf{Flow induced dissolution of femtoliter surface droplet arrays}
}
\\
Lei Bao\ttss{a}, Vamsi Spandan\ttss{b}, Yantao Yang\ttss{c}, Brendan Dyett\ttss{a},  Roberto Verzicco\ttss{d,b}, Detlef Lohse\ttss{b,e}, Xuehua Zhang\ttss{f,a,b*}

{\footnotesize \it
\noindent \ttss{a}Soft Matter $\&$ Interfaces Group, School of Engineering, RMIT University, Melbourne, VIC 3001, Australia. \\
\ttss{b}Physics of Fluids group, Department of Applied Physics, Mesa+ Institute, J. M. Burgers Centre for Fluid Dynamics $\&$ Max Planck Center Twente for Complex Fluid Dynamics, University of Twente, P.O. Box 217, 7500 AE Enschede, Netherlands. \\
\ttss{c}SKLTCS and Department of Mechanics and Engineering Science, College of Engineering, Peking University, Beijing 100871, China. \\
\ttss{d}University of Rome 'Tor Vergata', Via del Politecnico, Rome 00133, Italy. \\
\ttss{e}Max Planck Institute for Dynamics and Self-Organization,Am Fa$\ss{}$berg 17, 37077, G\"ottingen, Germany. \\
\ttss{f}Department of Chemical and Materials Engineering, Faculty of Engineering, University of Alberta, Edmonton, Alberta T6G1H9, Canada.\\ E-mail: xuehua1@ualberta.ca
}
\end{center}

\hrule
\vspace{1.0cm}

\begin{center}
{\bf Abstract}
\end{center}

The dissolution of liquid nanodroplets is a crucial step in many applied processes, such as separation and dispersion in food industry, crystal formation of pharmaceutical products, concentrating and analysis in medical diagnosis, and drug delivery in aerosols. In this work, using both experiments and numerical simulations, we \textit{quantitatively} study the dissolution dynamics of femtoliter surface droplets in a highly ordered array under a uniform flow. Our results show that the dissolution of femoliter droplets strongly depends on their spatial positions relative to the flow direction, drop-to-drop spacing in the array, and the imposed flow rate. 
In some particular case, the droplet at the edge of the array can dissolve about 30\% faster than the ones located near the centre. The dissolution rate of the droplet increases by 60\% as the inter-droplet spacing is increased from 2.5 $\mu$m to 20 $\mu$m. Moreover, the droplets close to the front of flow commence to shrink earlier than those droplets in the center of the array. The average dissolution rate is faster for faster flow. As a result, the dissolution time $T_{i}$ decreases with the Reynolds number Re of the flow as $T_{i}\propto Re^{-3/4} $. The experimental results are in good agreement with numerical simulations where the advection-diffusion equation for the concentration field is solved and the concentration gradient on the surface of the drop is computed. The findings suggest potential approaches to manipulate nanodroplet sizes in droplet arrays simply by dissolution controlled by an external flow. The obtained droplets with varying curvatures may serve as templates for generating multifocal microlens in one array.

\section{Introduction}

Microscopic droplet arrays on surfaces are key in many industrial, technological and analytic processes, because of their remarkable properties originating from their microscopic size, large surface-to-volume ratio, long time stability and defined spatial location.  These droplets play a crucial role in design of novel systems for chemical, catalytic and biological reactions  \cite{ANIE2010,Noji2005}, liquid-liquid extraction for separating or recycling harmful or valuable compounds  \cite{Chu2016,rezaee2006b}, or trace of anaytes in high throughput analysis for environment, forensic and biomedical monitoring \cite{Rissin2006,Shim2013}. They are also increasingly applied in drug screening \cite{Hatakeyama2006}, optimizing pharmaceutical products \cite{}, crystallization of proteins \cite{Benvenuti2007}, enzyme detection \cite{Jan2013,Rissin2006,Shim2013}, DNA sequencing \cite{Shimazaki2017}, cell culturing \cite{Zhao2017} and assembly of nanomaterials with multifunctionalities \cite{cai2008,lin2010}.

To be able to control and optimize the droplet-based processes, it is crucial
to quantitatively understand and control the droplet dynamics in presence of neighbouring droplets with high number density.  This is essential
and urgently needed not only for assembling colloidal particles to functional materials, for accumulating compounds in analysis with single molecular sensitivity \cite{Ota2012}, or producing drugs with most effect crystalline forms \cite{}, but also for 
advanced microfluidic systems as smart drug delivery platforms \cite{RIAHI2015} or as miniaturized chemical reactors with significantly reduced reaction time \cite{Weitz2011}. 
 These applications of droplets mainly rely on stability or desirable shrinkage rate of droplets.
 However, none of the systems studied so far have quantitatively revealed the dynamics of collective droplet dissolution in microfluidic channels, where femtoliter surface droplet arrays are exposed to a flowing surrounding phase. A quantitative understanding of collective effects associated with droplet spatial arrangement in an external flow is largely missing.

Recent advances in controlled droplet formation \cite{ZhangRMP2015,ZhangPNAS2015,Bao2015,lu2016} had made it possible to produce femtoliter surface droplets arrays with controllable size, volume and location on a microchannell wall. By the solvent exchange process, we have been able to form highly ordered surface droplet arrays with designed spatial arrangement on prepatterned substrates \cite{Bao2015}. Surface droplets nucleate and grow exclusively on hydrophobic domains in a constant contact angle mode followed by a constant contact area mode.  Such droplet arrays will serve as an ideal model system to quantitatively understand the dynamics in collective droplet dissolution under an external flow.

It is known that shrinking droplets on a substrate exhibit several modes, including constant contact angle \cite{Erbil2002}, constant contact area \cite{cazabat2010}, stick-jump \cite{debuisson2011c,debuisson2016}, or a mixed mode \cite{Zhang2015}. On a homogeneous smooth surface without chemical patterns, surface nanodroplets with polydispersed size dissolve in a mixed mode by decreasing both lateral diameter and contact angle. Droplets of the same initial size displayed individuality in dissolution rates, due to both pinning and collective effects \cite{Zhang2015,Xiaojue2018}. In analogy, droplet evaporation on arrays is influenced by convective flow driven by buoyancy \cite{Tol2005,Erbil2002}. An external flow enhances the dissolution rate of a single entrapped droplet \cite{Mustafa2016}, while natural convection plays a role in the dissolution of sub-millimeter drop in a stationary partially miscible liquid \cite{Laghezza2016,Dietrich2016}. The above situations are all different from the arrangement and dimensions of droplets to be studied in this work.

 In the present study, we experimentally and numerically study the dissolution of droplets in a square packed array. 
In particular, we look into how the droplet location and spacing influence their dissolution behavior and how the dissolution of droplets in an array is affected by the surrounding flow. 
Moreover, we demonstrate that controlled droplet dissolution can be potentially applied to create unconventional microlens-arrays with curvature gradients for micro-optical systems.  

This paper is organised as follows. In section 2, we describe the experimental technique and the framework for the numerical simulations. In section 3, we look at the effect of droplet position, spacing and imposed flow rate on the total dissolution time through both experiments and simulations. We conclude the paper with an outlook in section 4.

\section{Methods}

\begin{figure}
	\centering
	\includegraphics[scale=0.6]{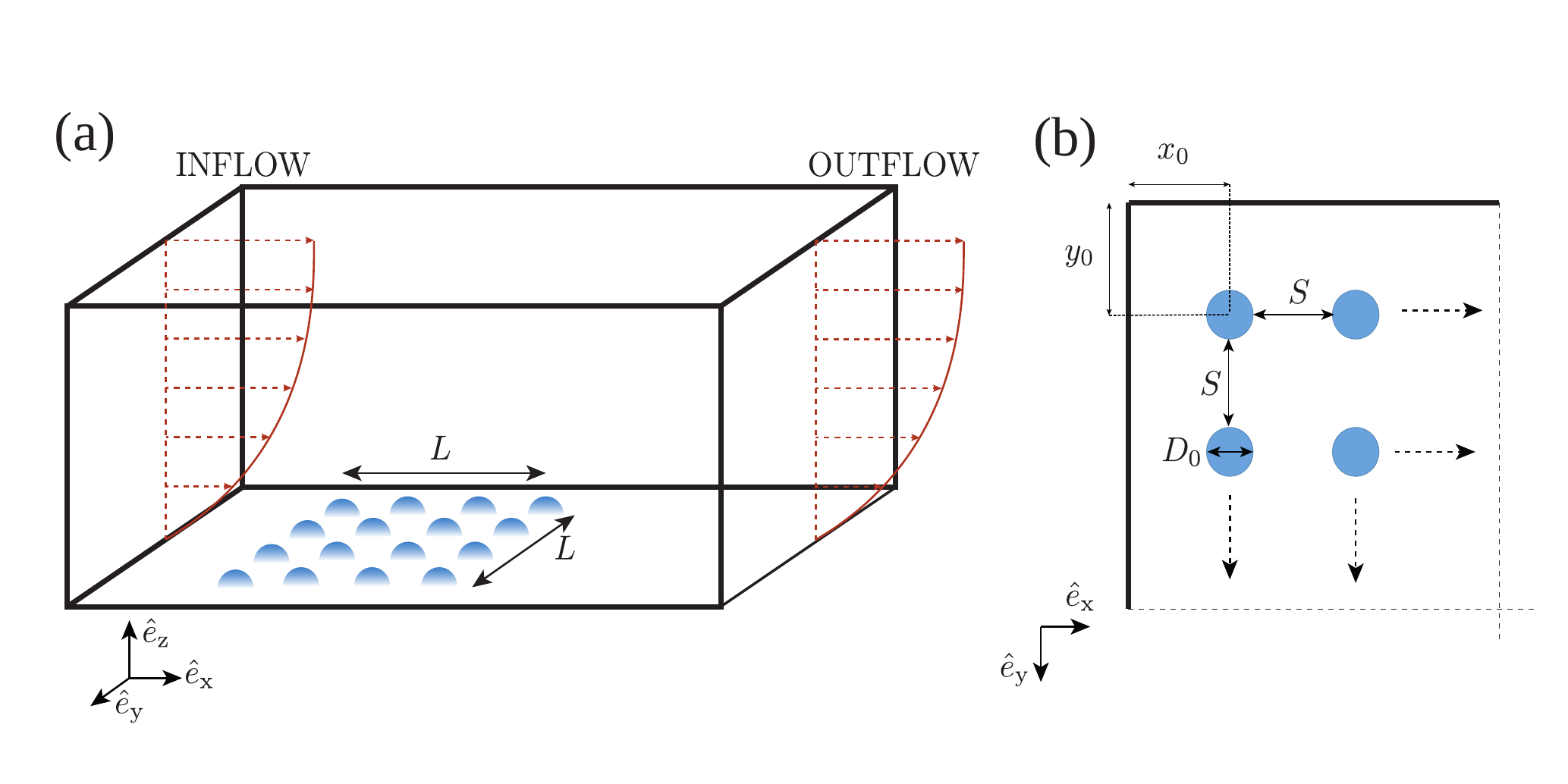}
	\caption{Schematics of the flow setup: (a) Arrangement of a droplet array of size $L$ in the microfluidic chamber with an imposed flow of fresh water. The chamber height, width and length in the $\hat e_\text z$, $\hat e_\text y$ and $\hat e_\text x$ directions are 0.63 mm, 15 mm and 63 mm, respectively. (b) Top view of the flow configuration: the initial drop diameter is $D_0$ and drops are equidistant from each other with spacing edge-to-edge $S$. Position of the corner drop in the first row is $(x_0,y_0$). The dotted arrows indicate a continuing array of drops in the $ \hat e_\text x$ and $\hat e_\text y$ directions, up to the array size $\textit{L}$.}
	\label{fig:schem}
\end{figure}

In Fig.\ref{fig:schem}(a), we show a schematics of the flow set-up while in Fig.\ref{fig:schem}(b) we show the top view along with the notations for the initial droplet lateral diameter ($D_0$) and the rim-to-rim inter-droplet spacing ($S$). $L$ is the distance from first droplet center to the last droplet center in the stream-wise ($\hat e_x$) and span-wise ($\hat e_y$) directions. 
The position of any drop in the array is normalised using $\tilde x = (x-x_0)/L$ and $\tilde y=(y-y_0)/L$, where $x_0$ and $y_0$ is the center position of the drop sitting at the corner in the first row as shown in Fig. \ref{fig:schem}(b).

\subsection{Experiments}

Highly ordered droplet arrays are prepared by performing solvent exchange on chemically patterned substrates \cite{Bao2015} where ethylene glycol dimethacrylate (EGDMA, 98\%, Sigma-Aldrich) is used as oil liquid for forming droplets. In brief, solution A containing 10\% EGDMA in 50\% ethanol-water solution liquid is first filled into a microfluidic chamber. It is then replaced by water saturated with EGDMA at a flow rate of 200 $\mu$L/min. During this solvent exchange, EGDMA droplets nucleate and grow within the hydrophobic domains on the pre-patterned substrates. The number of droplets, $D_{0}$ and $S$ is determined by the number, size and spacing of circular hydrophobic domains on the substrate.

During dissolution, water (18.2 M$\Omega$$\cdot$cm, Merck Millipore) is injected into the microfluidic chamber with a specific flow rate controlled by a syringe pump. The flow direction is $\hat e_x$ (stream-wise) as shown in Fig.\ref{fig:schem}. Total internal reflection fluorescence (TIRF) microscopy (Nikon N-Storm super resolution confocal microscope)
is used to gain high resolution images of individual dissolving droplets from the bottom view. In this case, 2 $\mu$M Nile red ($\geq$98\%, Sigma-Aldrich) is added into solution A for making the droplets. Reflection mode optical microscopy (Huvitz HRM-300) with 50$\times$ long focus lens is used to monitor the dissolution process of droplet arrays from the top view. The optical resolution is around 1.6 $\mu$m. The shape of droplets during the dissolution can be preserved by photo-polymerization. The morphologies of polymerized droplets are then characterized by atomic force microscopy (AFM, Dimension Icon-Bruker). While forming gradient microstructures, glass substrates with pre-patterned surface with 40$\times$40 domains are used. After a fully formed EGDMA droplet array, the water containing 0.5\% 2-hydroxy-2-methylpropiophenone (97\%, Sigma-Aldrich) is injected into the microfluidic chamber. The gradient microstructures obtained after partially dissolving the droplet arrays are photo-polymerized by placing the microfluidic chamber under a UV lamp (365 nm, 20 W) for about 15 mins. The diffraction patterns of the gradient microstructures under the white light illumination are then captured by a colour camera. 

\subsection{Numerical simulations}

We use direct numerical simulations (DNS) to solve the Navier-Stokes equations in a three dimensional Cartesian domain, using a second-order accurate finite-difference scheme with fractional time stepping. 
The flow solver is coupled with a versatile moving least squares (MLS) based immersed boundary method (IBM) which enforces the interfacial boundary condition of the immersed drops onto the flow. 
Additionally, an advection-diffusion type equation is solved for the concentration field. 
The boundary conditions for the flow setup are as follows: (i) inflow-outflow in the stream-wise $\hat e_\text x$ direction, (ii) no-slip walls in the wall-normal $\hat e_\text z$ direction, and (iii) periodic in the span-wise $\hat e_\text y$ direction. Hemispherical drops, which are discretised using triangular elements, are placed equidistantly from each other and on the bottom wall as shown in the schematic in Fig.\ref{fig:schem}. Through IBM, a free-slip boundary condition is imposed at the interface of the immersed drop and the incoming flow. The concentration field on the surface of the immersed hemispherical drop is kept fixed at saturation throughout the simulation. At any given instant, the concentration gradient on the surface of the immersed drop is computed by interpolating the concentration at the end-point of a numerical probe normal to every triangular element on the drop surface. The net mass flux from the drop into the flow is computed by summing up the flux from individual triangular elements belonging to an individual drop and the volume of the drop is reduced accordingly. In the simulations, a total of $7\times7=49$ drops are initialised and allowed to dissolve with an imposed flow rate. It is assumed that the centre of the contact patch of the droplets is fixed and they dissolve in a constant contact angle (CA) mode, i.e. the contact angle of the meniscus in contact with the bottom surface does not change with time. 

The droplet density ($\rho$) is 1051 $kg/m^{3}$, oil solubility in water ($C_{s}$) is 1.086 $kg/m^{3}$ and the diffusion constant is 9.36 $\times$ $10^{-10}$ $m^{2}/s$, all as in experiments. The grid resolution for the flow is 720 $\times$ 240 $\times$ 240 in the stream-wise, span-wise and wall-normal direction, respectively. The immersed drops are initialised with 2592 Lagrangian markers while local adaptive refinement is used as the drops shrink and become smaller. The simulations were run on 48 parallel processors. The Reynolds number of the flow based on the channel height and inflow velocity is set to 0.1. 
Additional details on the numerical techniques and validation of the code can be found in references \cite{ostilla2015multiple,spandan2017parallel}.

\section{Results and discussion}

\subsection{Two dissolution modes of individual droplet} 

\begin{figure}
	\centering
	\includegraphics[scale=0.6]{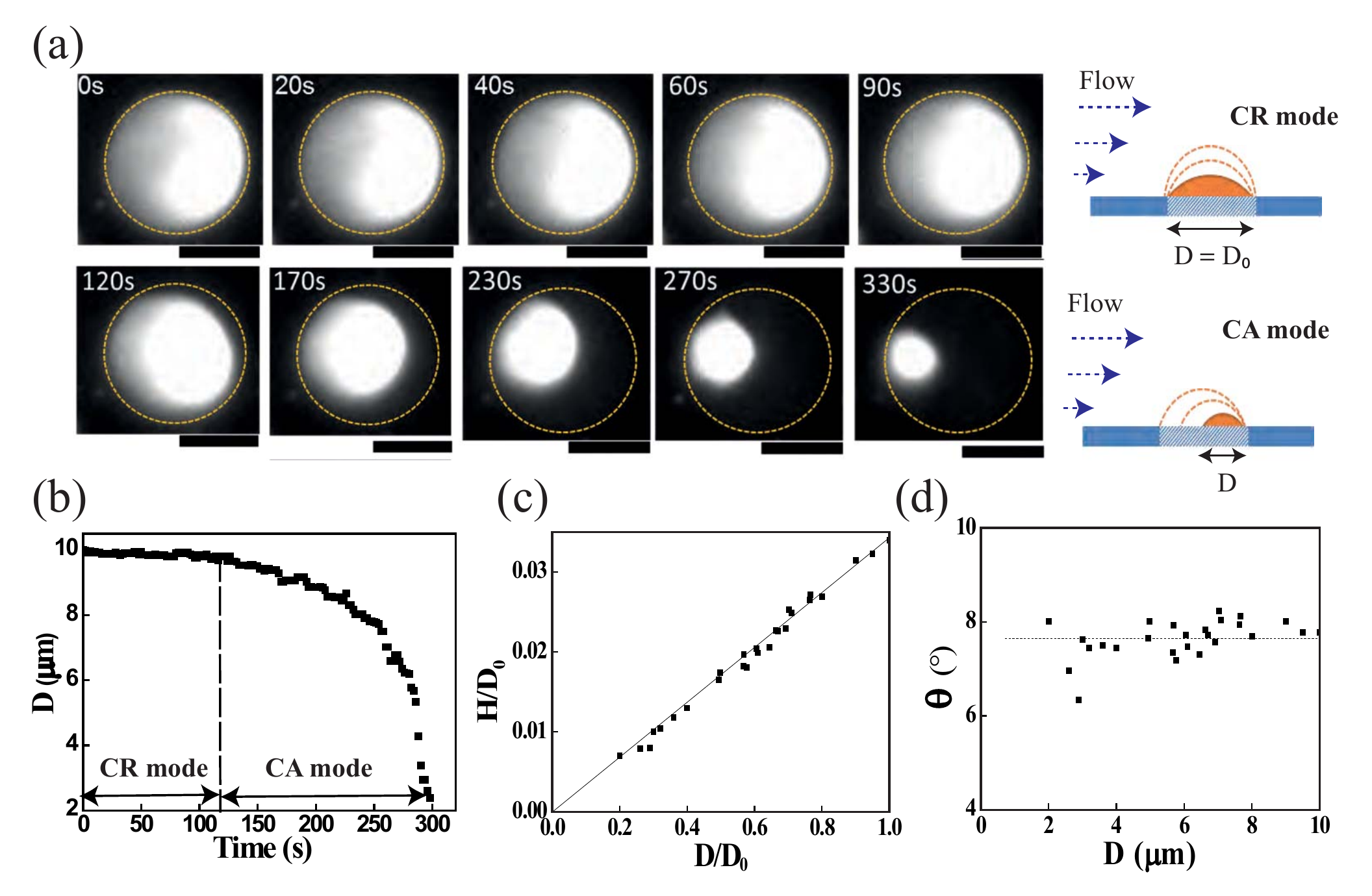}
	\caption{(a) Time evolution of a single droplet dissolution monitored by the TIRF confocal microscopy. The black scale bar is length of 5 $\mu$m. The sketches show corresponding dissolution modes. Top: constant contact area mode (CR mode); Bottom: constant contact angle mode (CA mode). 
	(b) Lateral diameter $D$ shrinks as function of time during the dissolution. 
	We note that due to optical resolution limit of the microscope for monitoring the droplets in-situ, the droplets are smaller than 1.6 $\mu$m could not be resolved. 
	(c) Height $H$ of the droplets as function of lateral diameter $D$ during the dissolution under CA mode. The height and lateral diameter are both normalised using $D_0$, which is the initial lateral diameter of the drops. (d) The contact angle $\theta$ hardly changes with the lateral diameter $D$ under CA mode dissolution.}
	\label{fig:2}
\end{figure}

In Fig.\ref{fig:2}(a), we show how a single droplet with $D_{0}$= 10 $\mu$m shrinks within a confined area in a flow of pure water at the rate of 200 $\mu$L/min. Up to 90 s, little change is observed in the lateral size of the droplets, attributed to be a constant contact radius (CR) dissolution mode. After 120 s, it is apparent
that the lateral diameter of the droplet decreases, suggesting a constant
contact angle (CA) dissolution mode. 
The combination between the pinning force on the micropattern boundary and the contact angle of the droplets determine the diameter for the transition of the model. \cite{Peng2017}
	
The other noticeable feature is that the three-phase contact line is pined at one side in CA mode dissolution. This phenomenon suggests that heterogeneities at the boundary of the hydrophobic-hydrophilic region induce strong pinning effects \cite{Zhang2015}. As a consequence, the center of droplet shifted as well. 

The lateral diameter as function of time is shown in Fig.\ref{fig:2}(b). In the stage of CA dissolution mode, the dissolution curve shows that the droplet diameter reduces laterally in a steady way but in a sharp way after it reaches to 8 $\mu$m. The shape of the droplet during the CA dissolution mode can be experimentally preserved by photo-polymerization and further imaged by AFM. Accordingly, we can plot the height of polymerized droplet as a function of its lateral diameter (Fig.\ref{fig:2}(c)). The linear relationship between the lateral diameter and the height confirms the CA dissolution mode. The corresponding contact angle is calculated to be approximately $8.0^ \text o$ as indicated in (Fig. \ref{fig:2}(d)). 

The focus in this work is the regime where droplet shrinks laterally, i.e CA dissolution mode. We then define the corresponding starting time of CA dissolution mode as $t=0$ for the following quantitative analysis.  
 



\subsection{Location dependence of droplet dissolution}

\begin{figure}
	\centering
	\includegraphics[scale=0.6]{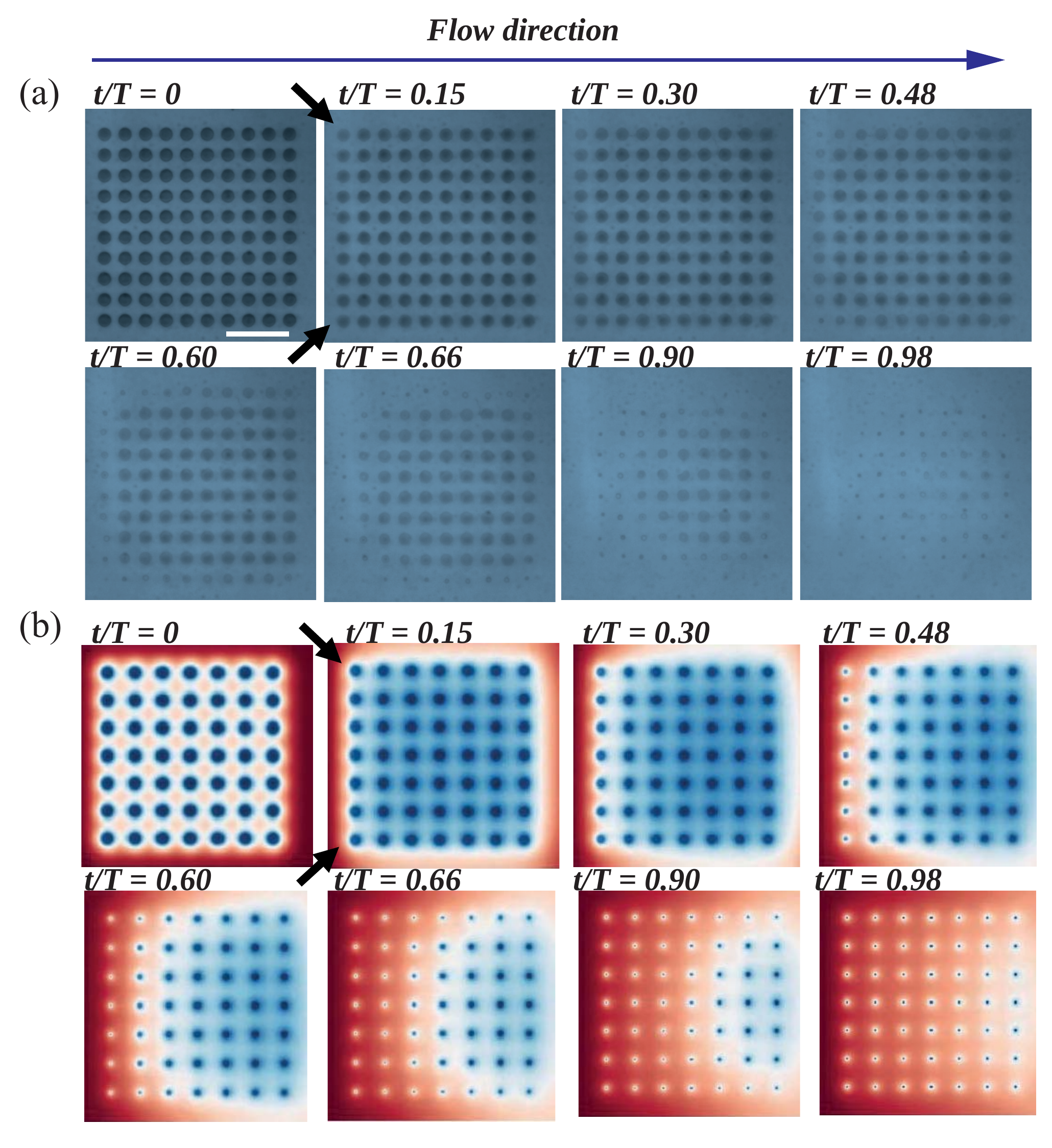}
	\caption{Time evolution of droplet dissolution: (a) Experimental results for a 10 $\times$ 10 droplet array with a spacing of S = 5 $\mu$m for a flow rate of Q = 200 $\mu L/min$. The length of scale bar is 45 $\mu$m. (b) Pseudocolour plot of the concentration field from numerical simulations for a 7 $\times$ 7 droplet array for corresponding flow conditions. Blue and red corresponds to oil and water, respectively. The water is injected into the chamber along $ \hat e_\text x$ direction. The black arrows indicate two corner droplet dissolving faster than other drops. Time $t$ is normalised using $T$, which is the total time for the droplet array to completely dissolve.}
	\label{fig:snap}
\end{figure}

We will now show the influence of the droplet position in an array on the  dissolution behavior in the CA mode. The representative experimental snapshots (Fig. \ref{fig:snap}(a)) show the overall dissolution process of droplets located in a square packed array. The initial lateral diameter of the drops is $D_0$ =10 $\mu$m while the spacing between drops is S = 5 $\mu$m, i.e. $S = 0.5D_0$.  
The dissolution initially occurs on the droplets located at the very corner of the array (i.e $(\tilde x, \tilde y)=(0,0)$ and $(\tilde x, \tilde y)=(0,1)$)), as indicated by black arrows. Strong dissolution effects are first shown at the outer drops and only later for the inner droplets of the array. It is apparent that dissolution is delayed along the stream-wise ($\hat e_x$) direction. The outer droplets in the array act as a shield to the inner droplets.  In Fig.\ref{fig:snap}(b), we show a pseudocolour plot of the concentration field obtained from the numerical simulations. It can be seen that the main features of the flow are very well captured. 

\begin{figure}
    \centering
	\includegraphics[scale=0.6]{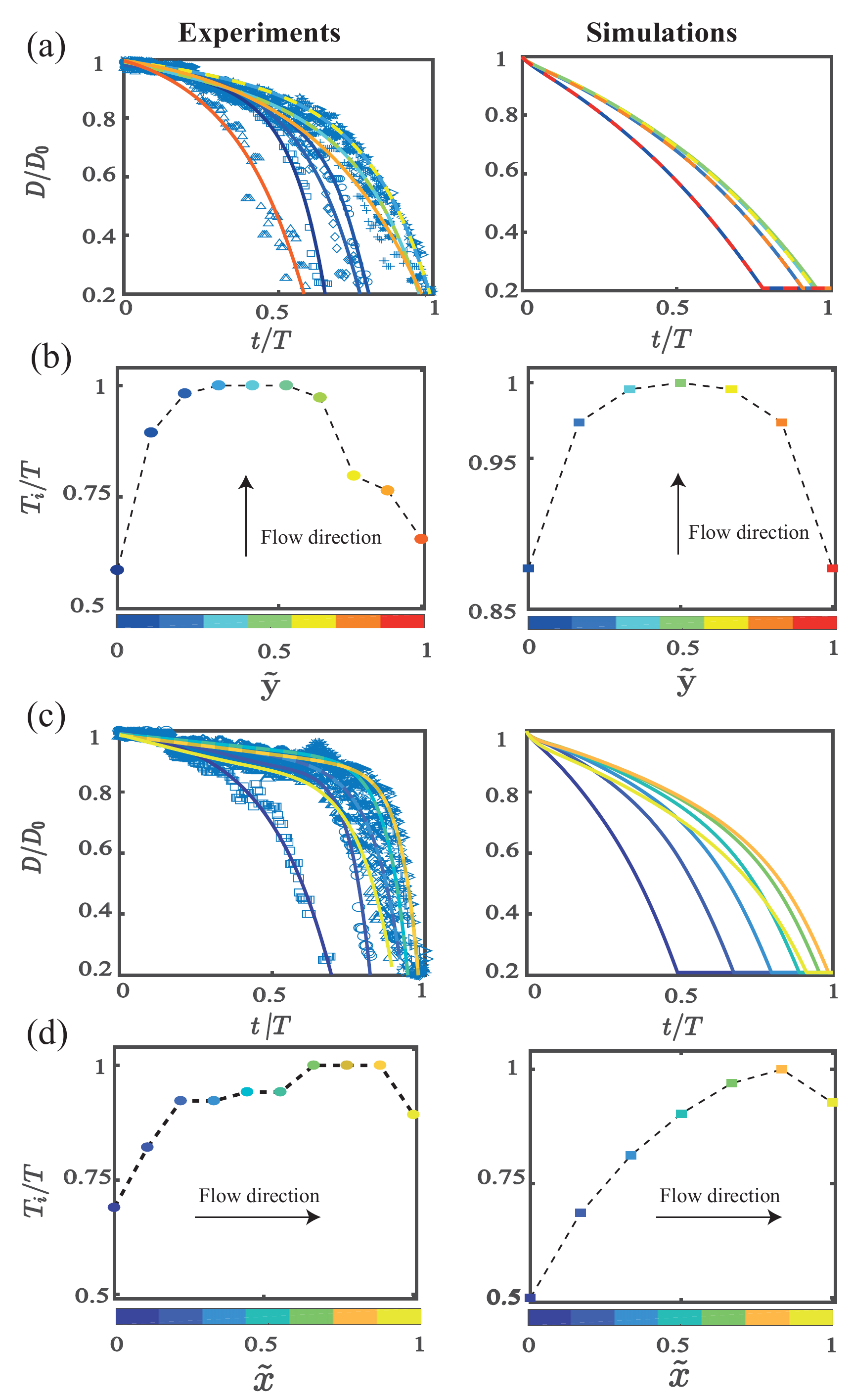}
	\caption{Time evolution of individual drop diameters in CA mode from experiments (left column) and numerical simulations (right column) along (a) the span-wise $\hat e_\text y$ and (c) stream-wise $\hat e_\text x$ directions. Total dissolution time ($T_i$) of droplets along the (b) span-wise and (d) stream-wise directions. In (a) and (b), $\tilde x=0$, i.e. the droplets are located at the first row and face fresh incoming flow, while in (c) and (d) $\tilde y=0.5$, i.e. it is the central column along the stream wise direction. $T_i$ is normalized using $T=\text {max}(T_i)$. The colour bars represent the location of the droplets along $\hat e_\text y$ direction for (a) and (b)  and $\hat e_\text x$ direction for (c) and (d).}
	\label{fig:4pp}
\end{figure}

To quantify the influence of the spatial position of the drop on its dissolution behaviour, we analyse the dissolution of drops located along specific rows and columns, respectively. In Fig.\ref{fig:4pp}(a), we plot the lateral diameter versus time from both experiments and simulations of droplets located in the first column, i.e. where the droplets are first hit by the front of the flow. 

The experimental measurements displayed in the left panel of Fig.\ref{fig:4pp}(a) clearly show that the droplets at $(\tilde x, \tilde y)=(0,0)$ and $(\tilde x, \tilde y)=(0,1)$ experience sharp decrease in the lateral diameter, while the rest of the drops dissolve more slowly. This feature is also confirmed by numerical simulations (right panel of Fig.\ref{fig:4pp}(a)), in which the corner drops also dissolve faster than the others. To quantify this, in Fig.\ref{fig:4pp}(b), we plot the total dissolution time ($T_i$) of individual drops along the first column ($\tilde x=0$) which clearly shows that the dissolution for the corner droplets is about 1.5 times faster than that of other droplets. We ascribe such difference to the different local concentration profile, determined by the neighbours of the corresponding droplet, i.e. collective effects. In the present case, each corner droplet has two close neighbouring droplets whereas each inner droplet is surrounded by three close neighbouring droplets. The additional nearby droplets lead to a higher saturation of the surrounding, which results in a reduced dissolution and thus an enhanced dissolution time of the inner droplets. The overall profile displayed in Fig.\ref{fig:4pp}(b) also reflects that the concentration gradient along the span-wise direction of flow changes in a parabolic shape. Compared to the simulated profile in the right panel, the asymmetry in the experimental profile is probably due to some distortion of the flow along the mid $\hat e_x - \hat e_z$ plane. 

As for droplets along the stream-wise direction, Fig.\ref{fig:4pp}(c) illustrates the diffusive loss of the droplets located in the central row with $\tilde y=5$. 
It is noticeable from both experiments and simulations, that in general the dissolution of droplets gradually slows down along the stream-wise ($\tilde x$) position.  The total dissolution time of corresponding droplets is summarized in Fig.\ref{fig:4pp}(d). The experimental normalized dissolution time has been extended from 0.7 to 1, which is comparable to the normalized simulated time from 0.5 to 1. The quantitative differences observed between the simulations and experiments mainly arise from the assumptions involving pinning, center-shift and dissolution of the drops undertaken in the simulations. The delayed dissolution along the flow direction indicates that the local concentration gradient for droplet downstream has significantly been reduced by the dissolving upstream neighbours \cite{Laghezza2016}.

However, there is one exception observed from both the experimental measurements and numerical simulations: The last droplet dissolves much faster than the droplets upstream of it. The reason is that the last droplet does not have any drop residing in its wake, which leads to a relatively higher concentration gradient in its wake. This leads to faster dissolution of the last drop or, in other words, a shorter dissolution time. One can 
expect that when the drops are placed far away from each other, this effect should be reduced and this is indeed observed when we study the effect of inter-droplet spacing on the dissolution behaviour later on. Despite this, the dissolution time of the last droplet is still significantly longer than that of first droplet, showing the key role of the flow direction.

\subsection{Inter-droplet spacing}

\begin{figure}
	\centering
	\includegraphics[scale=0.6]{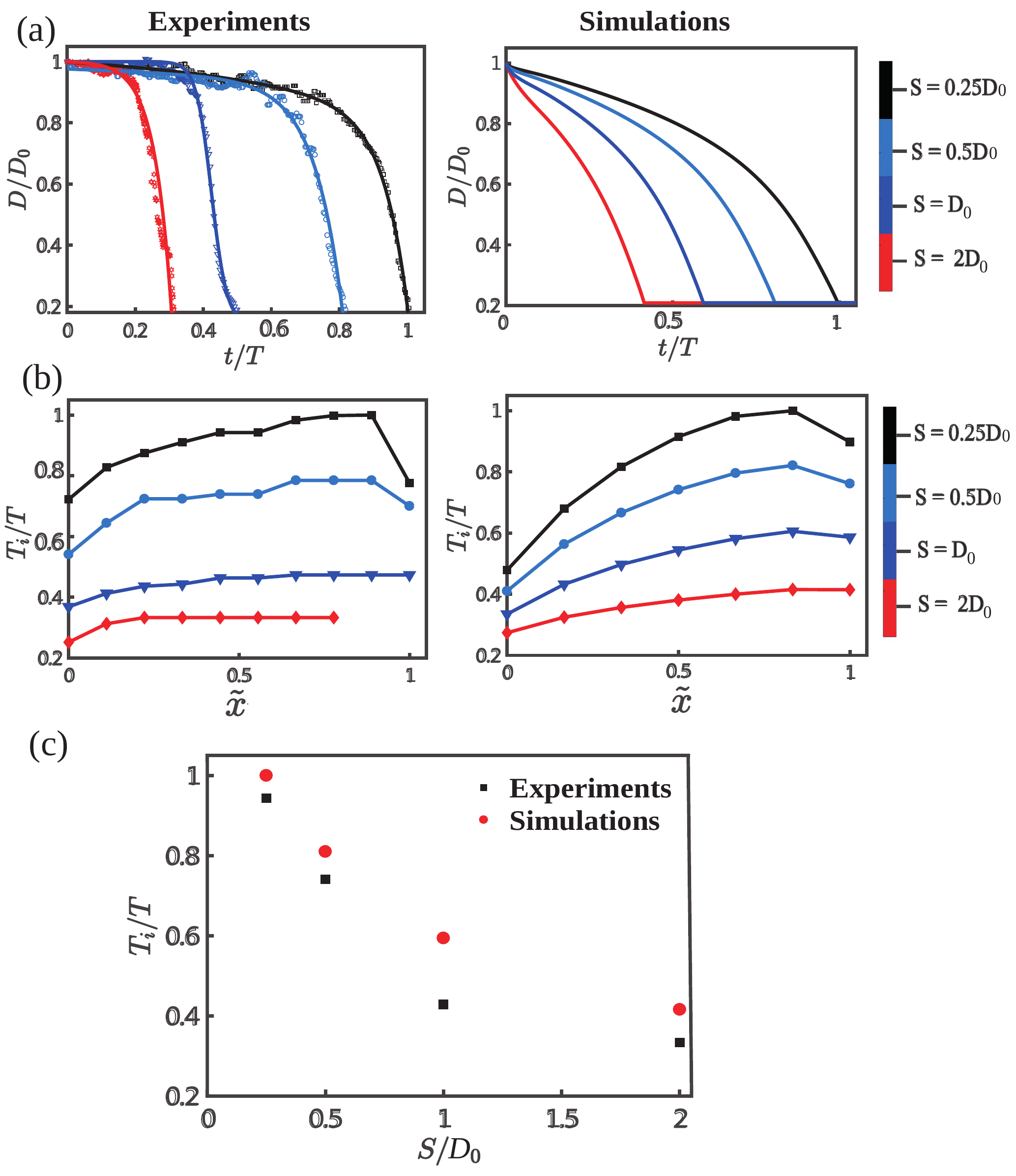}
	\caption{Droplet dissolution depends on the inter-droplet spacing $S$: (a) Time evolution of the central drop (i.e. $(\tilde x, \tilde y)=(0.5, 0.5)$) lateral diameter in CA mode for different inter-droplet spacings. (b) Total dissolution time ($T_i$) of droplets along the stream-wise direction. $T_i$ is normalized using $T=\text {max}(T_i)$. Note: Due to the limitation of optical window, only first seven droplets in one row can be captured in the case of $S=2D_0$. (c) The relationship of  $T_{i}$ with $S$. $T_{i}$ is normalised by the dissolution time (T) for the spacing of $S=0.25D_0$ and $S$ is normalised by $D_0$.}
	\label{fig:5pp}
\end{figure}

Next, we examine how the inter-droplet spacing ($S$) affects the overall dissolution behaviour. In total, four different arrays are prepared with an inter drop spacing $S$ of 2.5 $\mu$m, 5.0 $\mu$m, 10 $\mu$m, and 20 $\mu$m. Under an imposed flow of water at 200 $\mu$L/min, droplets with the same relative position ($\tilde x, \tilde y$) have different dissolution rates. The representative dissolution behaviour of the droplet situated in the centre of the array with different $S$ is shown in Fig.\ref{fig:5pp}(a). It can be immediately observed from both experiments and simulations, that the center droplet (i.e $(\tilde x, \tilde y)=(0.5,0.5)$) with the largest spacing $S=2D_{0}$ has a sharper dissolution slope than in the closely spaced array. Notably, the dissolution delayed along the flow direction has also been influenced, and Fig.\ref{fig:5pp}(b) summarizes $T_{i}$ of droplets in the central column with the corresponding $S$. The following features are revealed from both experimental and numerical results: (i) The lifetime of the droplets is significantly decreased as the spacing is increased due to better accessibility of the flow; (ii) The lifetime of the last droplet is extended only in closely packed array (i.e., $S=0.25D_{0 }$ and $S=0.5D_{0}$); (iii) The sequential dissolution along the flow direction is more pronounced in the closely packed array. All the evidences here quantify that the stronger collective effects occur in closely packed array as compared to loosely packed array. 

To give an quantitative estimate of the dissolution times, the total dissolution time of the central droplet $T_{i}$ is plotted versus the spacing $S$ in Fig.\ref{fig:5pp}(c). The experimental data are in good agreement with numerical simulations, showing that the dissolution time decays along with increasing spacing between the droplets. The dissolution rate of the  droplet can increase up to 60\% as $S$ increases from 0.25$D_{0 }$ to 20$D_{0 }$. It is expected that with large spacing between the droplets, the wake and the incoming flow for every drop is relatively fresh in comparison to closely packed droplets. This leads to a higher local concentration gradient which in turn to enhance the dissolution rate and thus reduce the droplet lifetime.

\subsection{Influence of external flow rate}
\begin{figure*}
	\centering
	\includegraphics[scale=0.6]{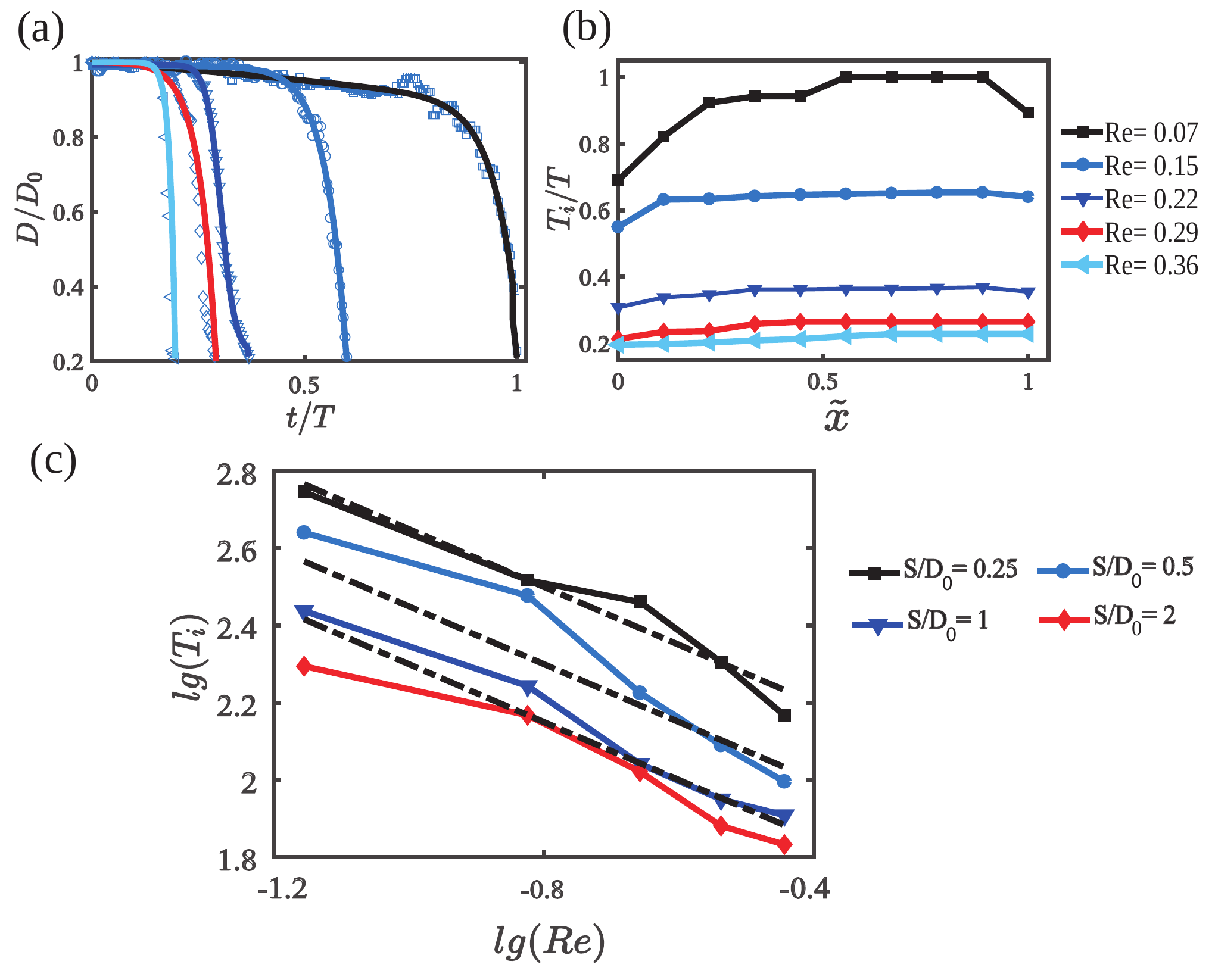}
	\caption{(a) Experimental dissolution curves for the central droplet $(\tilde x, \tilde y)=(0.5, 0.5)$ in CA mode under different flow rates, expressed in terms of Re for $S=0.5D_{0}$. (b) Total dissolution time ($T_i$) of droplets along the stream-wise direction under corresponding Re. $T_i$ is normalized using $T=\text {max}(T_i)$. (c) Experimental results of the total dissolution time $T_i$ of the central droplet with $(\tilde x, \tilde y)=(0.5, 0.5)$ as function of Re on a log-log plot for different droplet spacing $S$ as indicated. The dashed line shows the scaling law of  $T_i \propto Re^{-3/4}$.}
	\label{fig:6}
\end{figure*}

Finally, the rate Q of the fresh water flowing through the microchamber is varied. 
For different Reynolds number $Re = Q/(w\nu)$, the time evolution of the lateral diameter for the central droplets ($(\tilde x, \tilde y)=(0.5,0.5)$) is illustrated in Fig.\ref{fig:6}(a) as a representative case. Overall, the total dissolution time decreases at higher Re. As mentioned earlier, the droplets along the stream-wise direction have showed delayed dissolution in the closely packed pattern with $S = 0.5D_0$ under Re = 0.07 and Re = 0.15. (Fig.\ref{fig:6}b, black and blue curve). When the Re increases to 0.22 , each drop starts to shrink simultaneously without dependence on droplet location (Fig.\ref{fig:6}b), suggesting that the reduction of local concentrate gradient in the droplet array is more homogeneous by a faster external flow. Such feature is the evidence that faster external flow becomes dominant for droplet array dissolution, which has smeared out the local collective effect. Additionally, for a given array, the dissolution time $T_{i}$ scales as $T_{i}$ $\propto Re^{-3/4}$ (Fig.\ref{fig:6} c). This scaling reflects the laminar flow of Prandl-Blasius-Pohlhausen-type in the system \cite{Schlichting2000}.

\subsection{Microlens arrays with gradient structure}
 Before concluding the paper, we will demonstrate a potential application of the droplet array dissolution, namely in the fabrication of microlens arrays with curvature gradient. Microlens-gradient-arrays have superior optical properties for real-time 3D imaging \cite{Sara2012}, wide field optics \cite{Dumas2012}, and selective light harvesting. Owing to integrated different curvature unit lenses in one array, the focal plane of the microlens gradient array can be curved, largely overcoming the filed curvature aberration from conventional microlens arrays with flat focal plane \cite{Zhuang2014}. Such complicated lens array structure has been achieved by using sophisticate femtosecond laser direct writing technology \cite{Tian2015}. Herein, we propose a novel, simple and cost-effective method to generate microlens gradient arrays, simply based on droplet dissolution.   

\begin{figure}
	\centering
	\includegraphics[scale=0.6]{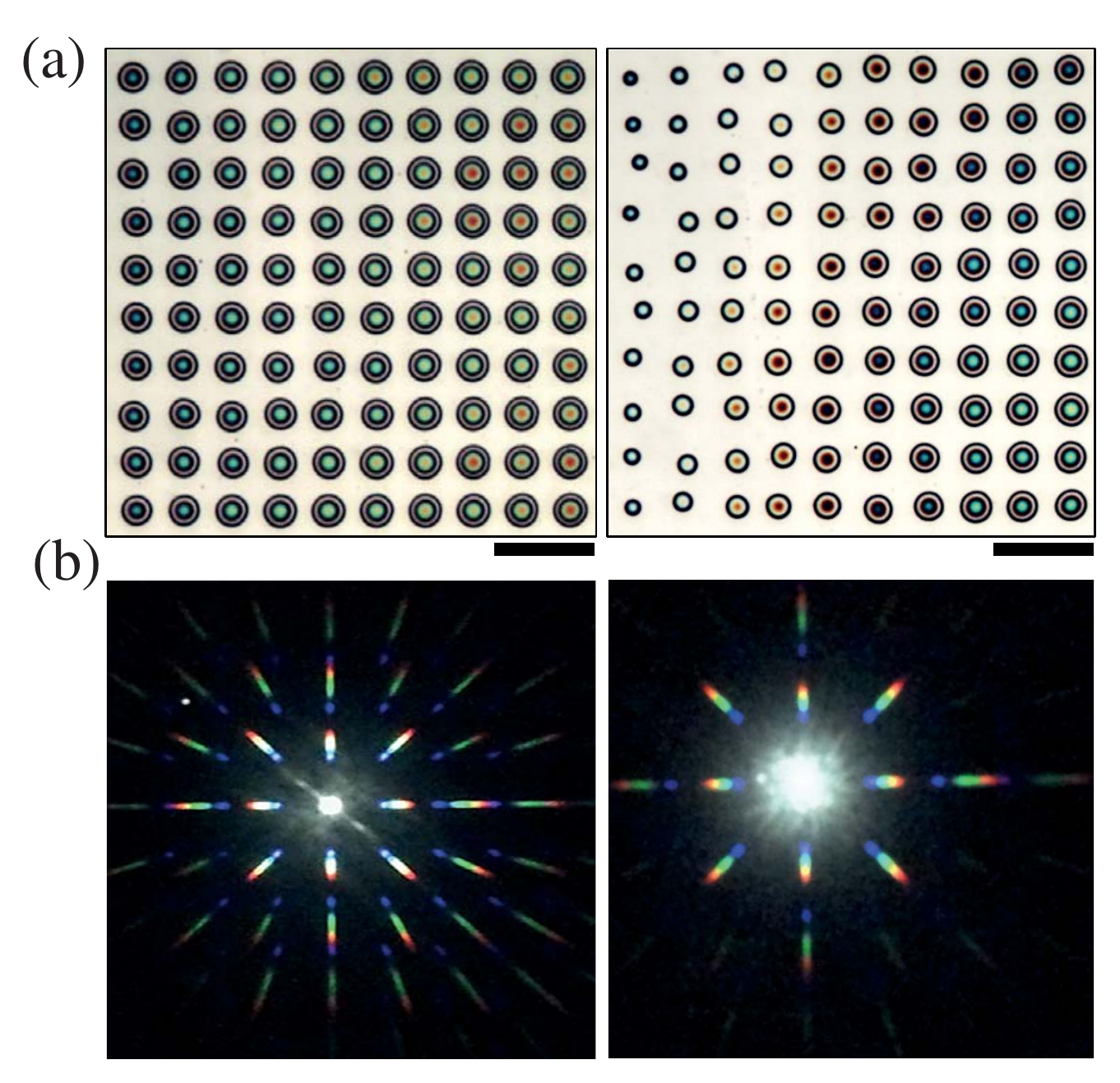}
	\caption{Gradient micro-structure based on droplet array dissolution: (a) Optical images of dissolved microlens arrays and (b) corresponding diffraction patterns. The scale bar is 30 $\mu$m.}
	\label{fig:7}
\end{figure}

As discussed in the earlier sections, because of strong collective and flow effect, the droplets on closely packed array at each location can experience different dissolution phases, which leads to various droplet curvatures on a single array. The corresponding morphology of the droplets has been well preserved by fast photopolymerization, resulting in microlenses on a array with gradient curvatures. Fig.\ref{fig:7}(a) shows the optical images of formed microlens arrays. The different colours and numbers of the Newton rings on each droplets clearly represent their different curvatures, which implies multifocus available on a single mircolens array. As a proof of concept, we test that the obtained arrays with microlens gradient display asymmetrical diffraction patterns under white light illumination (Fig.\ref{fig:7}(b)). The levels of diffraction order is increased along with the height or size of the microlenses which increases from left side to right side of the substrates. This simple demonstration shows the great potential to manipulate pathway of light selectively by microlens arrays with gradients, which will be significant to many optical applications. 

\section{Conclusions}
In summary, we have quantitatively demonstrated how the dissolution of femtoliter surface droplet-arrays depends on droplet location, spacing between them, and the imposed flow rate. The combined influence of collective effects and flow conditions on the lifetime of femtoliter droplet arrays have been quantitatively analysed via comparison of experimental and numerical results, which are quite consistent. Droplets surrounded with more neighboring drops take longer to dissolve, due to the collective effects as the droplets sense their neighbors through the surrounding flow field which is affected by the wake and the incoming flow. Such collective effects become more pronounced in a more closely packed array, in which the lifetime of droplets has been further extended. The collective effects are also reflected in dissolution rates of droplets in a row along the flow directions or across the flow direction; in the latter case the droplets at the edge dissolve faster. The overall dissolution is clearly accelerated by increasing flow rate in all cases, implying that the reduction of local concentrate gradient is enhanced by applying external flow. The immersed boundary method (IBM) code developed in this work will have predictive power for the dissolution of droplets with various initial sizes and spatial arrangements, far beyond what already covered in our experiments. 

Apart from our demonstrated fabrication of unconventional microlens arrays with optical advantages, the quantitative and in-depth understanding of the droplet dissolution in this work is highly valuable for many other droplet-based processes. 
In particular, the external flow and droplet spatial arrangement can be potentially manipulated to achieve uniform or gradually varying dissolution rate of individual droplets in the array. We envisage the gradually varied dissolution rates can be applied to effectively screen conditions for crystallisation of pharmaceutical products, micro-diagnostics, self-assembly of nanoparticles or microcompartmentalized chemical reactions. 

\section*{Acknowledgments}
 We thank Shantanu Maheshwari for support and discussions on the numerical simulations.
 L.B. acknowledges the support from RMIT Vice Chancellor Postdoctoral Fellowship. X.H.Z. acknowledges support from Australian
 Research Council (ARC, FT120100473). X.H.Z. and L. B. acknowledge support from ARC LP140100594. We also would like to thank Martin Klein Schaarsberg for
 assistance with data analysis, Dr Qiming Zhang for help on testing of optical properties and technical support from RMIT MicroNano Research Facility (MNRF). This work was supported by the Netherlands Center for Multiscale Catalytic Energy Conversion (MCEC), an NWO Gravitation programme funded by the Ministry of Education, Culture and Science of the government of the Netherlands. We acknowledge NWO for granting us computational time on Cartesius cluster from the Dutch Supercomputing Consortium SURFsara.
 
 \section*{Conflict of interest}
 There are no conflicts to declare.

\bibliographystyle{rsc}
\bibliography{rsc}


\newpage
\newpage

\end{document}